\documentclass[10pt,conference]{IEEEtran} 
\usepackage[utf8]{inputenc}

\IEEEoverridecommandlockouts

\usepackage{comment}
\usepackage{environ}     %
\usepackage{etoolbox}    %
\usepackage{graphicx}    %
\usepackage{amsmath,epsfig,epstopdf,amssymb}
\usepackage{adjustbox,lipsum}
\usepackage[hidelinks]{hyperref}
\usepackage{url}
\usepackage{tcolorbox}
\usepackage{adjustbox}
\usepackage{MnSymbol}
\newtheorem{definition}{Definition}

\newtheorem{example}{Example}

\usepackage{algpseudocode}%
\algnewcommand\algorithmicforeach{\textbf{for each}}
\algdef{S}[FOR]{ForEach}[1]{\algorithmicforeach\ #1\ \algorithmicdo}

\usepackage{float}
\floatstyle{boxed}
\newfloat{algorithm}{htbp}{loa}
\floatname{algorithm}{Algorithm}
\newfloat{algorithm}{htbp}{loa}
\restylefloat{figure}
\newfloat{figure}{h}{ext}
\floatname{figure}{Figure}

\usepackage{wrapfig}
\usepackage{multirow}

\usepackage{longtable}

\usepackage{multirow}
\usepackage{balance} %

\usepackage[backend=biber, abbreviate=true, dateabbrev=true, alldates=year, isbn=false, doi=false, urldate=comp, url=false, maxbibnames=9, backref=false, style=numeric-comp, language=american, giveninits=true, sortcites=true, sorting=none]{biblatex}
\AtEveryBibitem{\clearlist{location}} %
\AtEveryBibitem{\clearfield{series}} %
\AtEveryBibitem{\clearlist{language}} %

\usepackage{citation-cleaner}

\RemoveBibField{QiLAR15}{editor}
\RemoveBibField{YangZLT17}{editor}
\RemoveBibField{SPR}{editor}
\RemoveBibField{FraserA11}{editor}
\RemoveBibField{LinKCS17}{editor}

\usepackage[colorinlistoftodos,prependcaption,textsize=scriptsize]{todonotes} %

\newcommand{\head}[1]{\par\noindent\textbf{#1:}\space}

\title{Towards More Reliable Automated Program Repair by Integrating Static Analysis Techniques
\thanks{This work has been financially supported by the Research Council of Norway through the secureIT project (RCN contract \#288787).}
}

\author{\IEEEauthorblockN{Omar I. Al-Bataineh}
\IEEEauthorblockA{\textit{Simula Research Laboratory} \\
Oslo, Norway \\
omar@simula.no\vspace*{-2ex}}
\and
\IEEEauthorblockN{Anastasiia Grishina}
\IEEEauthorblockA{\textit{Simula Research Laboratory} \\
Oslo, Norway \\
anastasiia@simula.no\vspace*{-2ex}}
\and
\IEEEauthorblockN{Leon Moonen}
\IEEEauthorblockA{\textit{Simula Research Laboratory} \\
Oslo, Norway \\
leon.moonen@computer.org}\vspace*{-2ex}}

\begin{document}
\maketitle

\begin{abstract}
 A long-standing open challenge for automated program repair is the overfitting problem, 
 which is caused by having insufficient or incomplete specifications to validate whether a generated patch is correct or not. 
 Most available repair systems rely on weak specifications
 (i.e., specifications that are synthesized from test cases)
 which limits the quality of generated repairs.
 To strengthen specifications and improve the quality of repairs, 
 we propose to closer integrate static bug detection techniques with automated program repair.
 The integration combines automated program repair with static analysis techniques in such a way that bug detection patterns can be synthesized into specifications that the repair system can use.
 We explore the feasibility of such integration using two types of bugs: 
 arithmetic bugs, such as integer overflow, and logical bugs, such as termination bugs.
 As part of our analysis, we make several observations that help to improve patch generation for these classes of bugs. 
 Moreover, these observations assist with narrowing down the candidate patch search space, and inferring an effective search order.
\end{abstract}

\begin{IEEEkeywords}
automated program repair,
bug detection,
static analysis,
integer overflow,
non-termination,
conditional mutation.
\end{IEEEkeywords}

\section{Introduction}

\noindent
Automated program repair (APR) is an emerging research area that seeks to rectify bugs in programs 
by automatically generating patches that modify the source code~\cite{monperrus2018:automatic}.
Automated repair of bugs can significantly reduce the manual debugging effort, 
which is a very time-consuming and expensive activity in the software development process.
APR generally consists of four main steps: 
fault identification, fault localization, patch generation, and patch validation. 
In this paper, we focus mainly on \emph{patch validation},
in which the generated patch is extensively evaluated to ensure that the bug is resolved, 
and that the patch does not introduce any unwanted behavior.

APR needs a specification of correct program behavior to determine if a generated patch fixes the bug.
In the absence of having complete specifications for the programs they try to repair, 
existing APR approaches often resolve to using test cases as \emph{oracles} 
for determining if a patched program exhibits the desired (``correct'') behavior.
However, this means that generated repairs can only be considered as good as the test-suite itself:
one cannot be sure that the bug has been fixed for all possible inputs, 
and not just for the particular test cases.

The result is a long-standing open challenge for APR, known as \emph{patch overfitting}.
Patch overfitting is a condition where the patched program may pass the tests in the given test-suite, 
while it is failing for tests outside the test-suite. 
Since these patches are obtained by automated repair systems that rely on weak or incomplete specifications, 
there is no guarantee that the patches are general enough to address all possible inputs correctly.
Several solutions have been developed to alleviate the overfitting problem, 
such as symbolic specification inference~\cite{Nguyen,Mechtaev16}, 
machine learning-based prioritization of patches~\cite{Bader}, 
fuzzing-based test-suite augmentation~\cite{Gao}, 
and concolic path exploration~\cite{Shariffdeen21}. 
These solutions rely mainly on test cases and do not guarantee the general correctness of the patches.
They have in common that they do not \emph{solve} the overfitting problem, but aim to \emph{limit} its impact.

In search of alternative sources for specifications of correct behavior, 
we have identified static bug detection 
(also referred to as automated software inspection~\cite{anderson2003:tool}),
as a promising source for synthesizing accurate specifications that APR can use. 
Countless static analysis techniques have been developed to identify a wide variety of bugs,
such as division by zero, integer overflow, and out-of-bounds access. 
It is common for these techniques to employ formalized detection rules and patterns 
that capture the conditions under which certain types of bugs occur.
We argue that many of these bug detection patterns can be reformulated as specifications of correct behavior.
Exploiting this knowledge in APR can further alleviate, or even resolve, the overfitting problem,
improve the quality of repairs, and decrease the time spent searching for patches.

\head{Contributions} %
We propose an approach to address the overfitting problem of APR 
by using knowledge contained in static bug detection tools to enhance existing specifications. 
The approach synthesizes precise specifications for recurring classes of bugs
from the static analysis patterns and rules that are used to detect those bugs.
Moreover, it considers new classes of bugs whose fixes require the satisfaction of composite specifications.
We demonstrate the feasibility of the proposed approach using two classes of program bugs: 
(1) arithmetic bugs such as integer overflow, and (2) logical bugs such as termination bugs. 
We consider these bugs due to their widespread occurrence, 
and because solid static analysis tools are available for their detection and analysis.

The remainder of the paper discusses
the integration of static bug detection and APR at an abstract level in Section~\ref{sec:integrate},
followed by two motivating examples in Sections~\ref{sec:IOU} and~\ref{sec:terminate}.
Section~\ref{sec:implementation} sketches an initial prototype,
we discuss related work in Section~\ref{sec:relwork}, and we conclude in Section~\ref{sec:conc}.

\section{Integrating Static Bug Detection and Repair}
\label{sec:integrate}

\noindent
Bug detection is the natural step preceding (and succeeding) program repair. 
In today's APR approaches, bug detection is mainly done through testing, i.e., \emph{dynamic} program analysis, 
which leads to the challenges of using incomplete specifications and patch overfitting discussed before.
On the other hand, the literature on static program analysis for bug detection is rich and mature~\cite{dsilva2008:survey,bessey2010:few,sadowski2018:lessons}.
Many of these techniques build on automatically evaluated bug detection patterns and rules that describe the general conditions under which the bugs can occur, providing a systematic way to their detection.
We observe that many of these patterns can be captured using temporal logic or automata theory that can be interpreted as formal specifications of correct program behavior.

Therefore, one can take advantage of existing bug detection techniques to formulate accurate correctness specifications for recurring classes of bugs. 
We argue that combining static bug detection with APR is both possible and beneficial, 
as it will improve the overall reliability of the automated repair process.
The integration can be achieved formally
by defining APR as the process that generates the minimal patch which makes the program pass the given bug detection patterns.

We foresee several benefits that can be gained from integrating static bug detection with APR.
First, the integration will improve the reliability of repair systems by synthesizing accurate correctness specifications that do not solely rely on test cases,
but augment them with bug detection rules inferred using formal analysis techniques.
Second, by identifying the class of the bug through the application of bug detection tools, 
one can guide the repair engine to use specific program editing operators or specific repair strategies 
that have shown to be more promising for the particular class of the detected bug. 
This will considerably reduce the size of the patch space and the time required to generate repairs.
The integration of APR and static bug detection techniques can be performed in a variety of ways.
We describe two possibilities for integration:
\begin{enumerate}
 \item \emph{direct integration} by directly and repeatedly invoking the bug detection tool through the repair engine; 
 \item \emph{indirect integration} by extracting, collecting, and formalizing detection patterns for the bug being repaired. 
\end{enumerate}
The key advantage of the \emph{direct integration} approach is that it does not require a heavy implementation effort, and hence the integration may be performed in a straightforward manner.
However, the approach has several drawbacks:
(a) the tool needs to be invoked for each candidate repair.
This is a significant drawback, in particular if the size of the patch space is large: 
repeated calls of the detection tool would degrade the performance of the repair system and introduce considerable run-time overhead;
(b) every time the tool is invoked, it may generate information about all detected bugs in the program, which the user needs to examine  to extract the part that is relevant to the bug being repaired;
(c) bug detection tools are typically designed to support specific programming languages,
which imposes limitations on the applicability; 
and last but not least, (d) static analysis tools are known to suffer from false positives. 
These originate from the approximations that are needed because run-time values may be unknown at analysis time.
With direct integration, these false positives will propagate into the automated repair process.

The \emph{indirect integration} approach aims to collect and reuse bug detection patterns from bug detection tools. 
Manually deriving bug patterns from different tools is a tedious and time-consuming task.
It is therefore desirable to develop techniques that can extract (a.k.a., \emph{mine}) detection patterns from the tools,
and reformulate them as correctness specifications in a format that is acceptable by the repair tool. 
This is a challenging open problem to address in general.
However, bug detection tools increasingly do not hard-code their detection rules, 
but make them configurable and customizable through pattern- and query-languages.
As a result, it is possible to mine those rules automatically on a tool-by-tool basis.
Key advantages of the indirect integration approach are as follows: 
(a) correctness of patches is automatically guaranteed provided that the developed patterns are correct; 
(b) the effort is reusable, as formal rules can be developed once and reused in the future; 
and (c) the approach does not suffer from false positives since we build on the bug detection \emph{rules}, 
not on the static approximation of program values at run-time.

In the following two sections, we consider two example classes of bugs and show how direct and indirect integration can be applied to improve the overall quality of the repair process by synthesizing accurate specifications. 

\section{First Motivating Example: Arithmetic Bugs}
\label{sec:IOU}

\noindent
Arithmetic calculations affect a wide variety of software applications, 
including safety-critical systems such as control systems for vehicles, medical equipment, and industrial plants. 
In this section, we study arithmetic bugs, in particular integer overflow and integer underflow,
and show how one can extract and formulate detection rules for this class of bugs. 

\subsection{Arithmetic Bugs (Integer Overflow)}

\newcommand{\INTMAX}{\ensuremath{\text{\textsc{intmax}}}}
\newcommand{\INTMIN}{\ensuremath{\text{\textsc{intmin}}}}
\newcommand{\IO}{\ensuremath{\text{IO}}}
\newcommand{\IU}{\ensuremath{\text{IU}}}
\newcommand{\f}[1]{\ensuremath{\text{\textit{#1}}}}
\newcommand{\checkflow}{\f{isOverflow}}

\noindent
We discuss two classes of arithmetic bugs, namely integer overflow (\IO) and integer underflow (\IU).
\emph{Integer overflow} is a common bug that occurs when the computation of an arithmetic operation, 
such as multiplication or addition, exceeds the maximum size of the integer type used to store it.
An \IO{} condition may give results leading to unintended behavior that can compromise a program's security. 
To address this type of bugs, we can follow an indirect integration approach and extract relevant rules for the detection of IO bugs from the static analysis tool IntRepair~\cite{Paul2019}. 

Let $x$ and $y$ be integer variables, and \INTMAX{} be the positive upper bound that the variables can store, 
and \INTMIN{} be the negative lower bound that the variables can store.
The first rule considers an arithmetic expression that adds two integer variables:
\begin{equation*}
 \checkflow(x + y) = 
 \begin{cases}
 \IO , & \text{if~} (x > 0 \land y > \INTMAX - x) \\
 \IU, & \text{if~} (x < 0 \land y < \INTMIN - x) \\
 \f{false}, & \text{otherwise.} \\
 \end{cases}
\end{equation*}
The second rule considers subtraction of two numbers: 
\begin{equation*}
 \checkflow(x - y) = 
 \begin{cases}
 \IO , & \text{if~} (x > 0 \land y < x - \INTMAX) \\
 \IU, & \text{if~} (x < 0 \land y < \INTMIN - x) \\
 \f{false}, & \text{otherwise.} \\
 \end{cases}
\end{equation*}
The third rule considers multiplication of two numbers: 
\begin{equation*}
 \checkflow(x * y) = 
 \begin{cases}
 \IO , & \text{if~} (x > 0 \land y > \INTMAX / x) \\
 \IU, & \text{if~} (x < 0 \land y < \INTMIN / x) \\
 \f{false}, & \text{otherwise.} \\
 \end{cases}
\end{equation*}
It is easy to see that the extracted rules are sound, given the semantics of the basic arithmetic operators $\{+, -, *\}$.
Note that the rules are written in such a way that the conditions concern the individual variables, 
not the combined expression, to ensure that the preconditions of computing the expression are checked.
Moreover, the rules can be easily extended to expressions with a larger number of operands and operators, so that they can be applied to detect \IO{}/\IU{} bugs in more complex arithmetic expressions.
For example, to check flow in the expression $e = (a + b - c)$, 
we need to apply the two rules, $\checkflow(x + y)$ and $\checkflow(x - y)$. 
These rules will check the sub-expressions $\{a + b, b - c, a + b - c\}$.

 \subsection{Correctness Specifications for Repairing \IO{}/\IU{} Bugs}

\noindent
A fundamental challenge in APR comes from missing complete specifications of the intended program behavior. 
Since a complete specifications of correct behavior is usually not available, 
existing program repair techniques rely mainly on test cases. 
However, in the case of arithmetic expressions,
the limited domain allows for synthesis of a complete specification by examining the semantics of the expression. 

Let $\checkflow(e)$ be a bug detection rule that checks whether the expression $e$ can lead to integer overflow or underflow. 
Let also $\f{split}(e)$ be a function that splits the expression $e$ into its
basic sub-expressions while taking into account the order at which sub-expressions are executed. 
The function $\f{split}(e)$ uses basic arithmetic operators and left and right parenthesis as splitting delimiters when decomposing expressions into simpler ones. 
For example, let $e = (a + b * c)$.
Then $\f{split}(e) = \{b * c, a + d\}$, where $d = (b * c)$ and hence, 
we need to apply the rules $\checkflow(a+d)$ and $\checkflow(b*c)$ 
to check whether the expression $e$ is a bug-free expression.

A valid patch for an IO/IU buggy expression $e$ needs to satisfy a composite correctness property: 
(i) under all possible valuations of variables in $e$, 
none of them will store a value greater than the maximal allowed value, 
and (ii) the semantics of $e$ are preserved in the patch after being mutated.
This can be captured in a correctness specification as follows:
\begin{equation} \label{spec_io}
 \f{Spec}_\f{IO} = \forall_{e_s \in \f{split}(e')} (\checkflow(e_s) = \f{false}) \land e' \equiv e
\end{equation}
where $e'$ represents a mutated version of $e$.
The splitting of the expression $e'$ into its basic expressions is necessary 
to ensure that fixing a bug in some parts of the expression does not introduce new bugs in other parts of the expression.
Note that we do not consider IO bugs as semantic bugs but rather as memory allocation bugs or formulation bugs, 
so that feasible patches can be generated by simply reformulating or rewriting the expression. 
From specification $\f{Spec}_\f{IO}$, 
it follows that the mutation function that generates candidate repairs $e'$ for $e$, 
needs to meet the following requirement:
\begin{equation}\label{IOvalidity}
 \f{validity}(e') = 
 \begin{cases}
 \f{valid} & \text{if~} e' \equiv e \\
 \f{invalid} & \text{otherwise} \\
 \end{cases}
\end{equation}
One can employ contemporary SMT solvers to check the equivalence of two arithmetic expressions $e_i$ and $e_j$. 
This can be performed by checking the satisfiability of a formula of the form $\varphi = (e_i \neq e_j)$. 
If there exists an assignment to the variables of $e_i$ and $e_j$ that make $\varphi$ satisfiable, 
then the two expressions are not semantically equivalent. 
On the other hand, if the formula $\varphi$ is unsatisfiable, then $e_i$ and $e_j$ are semantically equivalent. 
Of course, the completeness of this validation approach for IO bugs is 
relative to the completeness of the SMT solver that is employed to check the equivalence.

\subsection{Repair Procedure for IO/IU Bugs} \label{sec:RepairIO}

\noindent
We discuss two possible ways to repair the class of IO bugs. 
First, certain IO bugs may be repaired by \emph{rewriting} the buggy arithmetic expression 
in such a way that its sub-expressions no longer cause an integer overflow.
Rewriting rules that transform arithmetic expressions into semantically equivalent expressions
can be extremely useful for dealing with IO bugs.
For example, consider the arithmetic expression $e = (a + b - c)$, with $a, b, c > 0$.
Since addition and subtraction have the same precedence and are left-associative, 
the following mutations for the expression $e$ can fix an IO bug in $a + b$ via rewriting 
$\f{mutants} (e) = \{ (a - c + b), (b - c + a) \}$, 
as long as certain constraints on the values of $a$, $b$, and $c$ are true.

Second, the IO bug may be repaired by using a \emph{variable widening} technique. 
The technique converts a variable from one data type into another data type that accepts a broader range of values.
However, care needs to be taken that such repairs do not violate the semantics of the original program. 
It is not enough to just widen the type of the variable that triggered the overflow. 
To ensure the overall correctness of the patched program and to avoid introducing new bugs to the program, 
we may need to widen all \emph{dependent} variables in the program, 
i.e., all variables that are defined using widened variables. 
Consider the statements $s_i: e = a + b - c$ and $s_j: d = e * f - g$. 
It is easy to see that the variable $d$ is a dependent variable whose value depends directly on the values of $e$, $f$, and $g$.  
Hence, if we widen the variable $e$, 
we may need to widen the variable $d$ as well, due to the dependency relationship between the two variables.  
To determine whether or not the variable $d$ needs to be widened, 
we check the statement $s_j$ against the developed IO detection rules while taking the new datatype of variable $e$ into account. 
The validation process of variable widening as a strategy to patch IO bugs is more complex than one might anticipate: 
it requires examining not only the buggy expression but also all other expressions to which the widened variables contribute.
Moreover, another source of potential bugs that needs examination are (external) library functions that consume variables which were widened as part of the repair, as they may be incompatible with the widened datatype.

With these caveats in mind, it is possible to develop a search-based repair approach 
that only considers feasible patches for IO bugs cf.\ Eq.~\ref{IOvalidity}. 
The candidate patches then need to be checked against the correctness specification $\f{Spec}_\f{IO}$. 
This approach does not rely on test cases for validating generated repairs, 
but on formal IO/IU bug detection rules instead.

\section{Second Motivating Example: Termination Bugs}
\label{sec:terminate}

\noindent
Proving termination of programs is a challenging and important problem, 
where even partial solutions can significantly improve software reliability and programmer productivity.
A huge body of work has been published on proving termination of programs based on a variety of techniques, 
such as abstract interpretation~\cite{Berdine2007, Chawdhary2008, TsitovichSWK11}, 
bounds analysis~\cite{Gulwani09, Gulwani09speed}, 
ranking functions~\cite{Bradley2005, Cousot05}, 
recurrence sets~\cite{Gupta2008, Harris2010}, 
and transition invariants~\cite{Kroening2010, Podelski2004}. 
The most popular technique to prove termination is through the synthesis of a ranking function, 
a mapping from the state space to a well-founded domain, 
whose value monotonically decreases as the computation progresses.
We will refer to the class of logical bugs in programs 
that lead to non-terminating loops as \emph{termination bugs}
and analyze the conditions under which they can be automatically repaired.

\subsection{Termination Bugs}

\noindent
Termination bugs have received relatively little attention in the APR literature.
Repairing termination bugs can be non-trivial for several reasons. 
First, program termination is undecidable in general. 
Second, fixing termination bugs requires not only ensuring termination of the program under repair, 
but also preserving the intended semantics of the program. 

Another key challenge when dealing with termination bugs is the difficulty of proving the presence of these classes of bugs using test cases. 
It is not self-evident how long one would need to run the program to prove non-termination.
As a result, current repair approaches that rely on test suites cannot validate generated patches for detected termination bugs, since the size of the program input space can be extremely large or even infinite. 
However, it is possible to compute the expected upper bound for termination of a given \emph{loop program}, i.e., a program containing a loop, 
based on an analysis that takes into account the structure of the loop program and the architecture of the computer that executes the program.
Termination provers can be employed to assist with this computationally complex task and prove non-termination in an automated manner. 

Before proceeding further, let us formally introduce some basic notions that we use throughout the paper, 
namely the notion of halting statements, termination bugs, and the termination repair problem.
\begin{definition} \label{HltStatements} \textbf{Halting statement}.
We refer to a reachable statement $s$ in a program $P$ as a \emph{halting statement}, %
iff $s$ meets one of the following conditions:
\begin{enumerate}
  \item $s$ is a special type statement whose execution causes termination, such as a RETURN statement, 
     and $s$ is not part of a function that is called by another function;
  \item $s$ is the final statement of $P$, or the final statement of a function that is not called by another function.
\end{enumerate}
\end{definition}
\begin{definition} \textbf{Termination bug}.
Let $P$ be a program containing a set of halting statements $H \subseteq P$ at which the program terminates (cf. Definition~\ref{HltStatements}).
We say that program $P$ contains a \emph{termination bug} iff there exists a set of inputs $i$ that prevent $P$ from reaching any of the halting statements $H$, regardless of how long the program is running.
\end{definition}
\begin{definition} \textbf{Termination repair}.
Let $I$ be the set of possible inputs to a buggy non-terminating (NT) program $P$. 
Let $\varphi$ be a property that captures the intended semantics of $P$. 
Then termination repair aims to synthesize a new program $P'$ 
that is (semantically) similar to the original buggy program $P$ 
such that for each set of inputs $i \in I$ the program $P'$ reaches some halting statement $s \in H$,
and $\f{run}(P', i) \models \varphi$ (i.e., the result of $P'$ on input $i$ satisfies the property $\varphi$). 
\end{definition}

One key issue that distinguishes termination bugs from other classes of bugs 
is that fixing termination bugs requires the program to satisfy a \emph{composite property}: 
a termination property and a functional property. 
That is, to fix a termination bug, one needs to ensure termination for each possible input (\emph{termination property}) 
while preserving the semantics of the program (\emph{functional property}). 
This further increases the complexity of the repair problem of termination bugs. 

There are several benefits of using termination provers in the process of repairing termination bugs.
First, termination provers can be used to prove the presence of termination bugs formally. 
Second, they can be used to check the soundness of generated patches in an automated way.
This helps to avoid the construction of complex proofs and the exhaustive exploration of the input space of the patched program.
Third, termination provers can provide information that can be used to automatically find counterexamples, 
which in turn can be used to guide the repair algorithm to generate valid repairs.

\subsection{Correctness Specifications for Repairing NT Loops}

\noindent
There are many different ways to solve termination of a given non-terminating (NT) loop program $P$, 
for example, by mutating the termination expression $\mathcal{C}(P)$ of the loop $P$,
or by mutating the set of expressions $E$ that affect the termination expression $\mathcal{C}(P)$.
However, some mutations of $\mathcal{C}(P)$ or $E$ may fix the termination bug, 
but break the intended semantics of the loop.
To satisfy the functional property, it is, therefore, necessary to fix termination while preserving the semantics of the loop. 
This composite requirement can be synthesized in a two-part specification as follows:
$$
\f{Spec}_\f{Term} = ( P' \models \varphi_\f{term} \land P' \models \varphi^P_\f{sem} )
$$
where $P'$ represents a mutated version of the non-terminating loop program $P$, 
$\varphi_\f{term}$ represents a property that ensures termination of $P'$,
and $\varphi^P_\f{sem}$ is a property that captures the intended semantics of $P$, which is then checked against $P'$. 
The termination property $\varphi_\f{term}$ can be synthesized from the termination patterns that are used by the termination tools like AProVE~\cite{AProveIJCAR14} and 2LS~\cite{cdksw2015} for the loop $P'$, 
or by simply running the termination provers directly against the mutated loop $P'$ to check termination. 
The functional property $\varphi^P_\f{sem}$ can be synthesized in a variety of ways. 
One way would be to use previously known passing or successful test cases to check whether the semantics of the program are preserved after deploying the patch. 

Satisfaction of the complete specification $\f{Spec}_\f{Term}$ ensures that the termination bug is fixed and that the semantics of the loop are preserved.
Consider, for example, a loop program that aims to sort an array in ascending order. 
Then the property $\varphi^P_\f{sem}$ checks whether the array is sorted correctly after termination, 
while the property $\varphi_\f{term}$ checks whether the loop will terminate after a finite number of iterations. 

Next, we need to define the process of validating generated patches for termination bugs.
Observe that due to the computational expense of employing termination provers, 
it is highly desirable to implement a 2-step process for validating potential patches for termination bugs.
In the first step, we prune the set of candidate patches using any available test cases to reject invalid patches for the program under repair, 
and in the second step, we formally check the validity of generated plausible patches by running termination provers.
This 2-step approach helps to considerably reduce the overhead introduced by running termination provers.
The passing test cases $T_p$ are used to model the expected correct behavior of the program. 
\begin{definition} \label{ValidityDef} \textbf{Validity of Patches for Termination Bugs}.
Let $P$ be a buggy non-terminating loop program and $T = (T_p \cup T_f)$ be a test-suite that consists of the set of passing test cases $T_p$ and the set of failing test cases $T_f$.
Let $P'$ be a candidate patched version of $P$ and $A$ be a termination prover that returns one of the following answers $\{\f{terminating}, \f{non-terminating}, \f{unknown}\}$. %
We say that the program $P'$ is a valid patched version of the non-terminating program $P$ iff all of the following conditions hold:
\begin{enumerate}
 \item all failing test cases in $T_f$ pass in $P'$, 
 \item none of the passing test cases in $T_p$ fail in $P'$,  
 \item the termination prover $A$ returns ``terminating'' when analyzing termination of the loop program $P'$. 
\end{enumerate}
\end{definition}
As mentioned earlier, the effectiveness of search-based ``generate-and-validate'' repair approaches can be disputed, 
because they typically cannot provide patch correctness guarantees. 
However, as we see here, the integrating of these techniques with solid bug detection techniques can significantly improve the effectiveness of the combined approach.
AProVE and 2LS are among the most reliable termination analysis tools. 
They take a program as input and return one of three answers: 
\emph{terminating} ($\f{TR}$), \emph{non-terminating} ($\f{NT}$), or \emph{unknown} ( $\f{UN}$).
In general, when the prover returns a definite answer for a given program
(i.e., $\f{answer} \in \{\f{TR}, \f{NT}\}$), 
the answer is with high confidence a valid answer.

AProVE is a system for automated termination and complexity proofs of term rewriting systems. 
2LS is a CPROVER-based framework that reduces program analysis problems expressed in second-order logic, 
such as invariant or ranking function inference, to synthesis problems over templates.
In the 5th Competition on Software Verification (SV-COMP'16), 
AProVE was the strongest tool for the termination category, 
while 2LS has been shown to be a powerful tool for proving termination for larger programs with thousands of lines of code~\cite{cdksw2015}. 
We use test cases together with termination provers to check the validity of generated patches as follows: 
 $$
 \small
 \f{validity} (p) = 
 \begin{cases}
 \f{valid}, & \text{if~} (\forall_{t \in T} (\f{p} \vdash t) \land A(\f{p}) = \f{TR}) \\
 \f{plausible}, & \text{if~} ((\forall_{t \in T} (\f{p} \vdash t) \land A(\f{p}) = \f{UN}) \\
 \f{invalid}, & \text{if~} ((\exists_{t \in T} (\f{p} \nvdash t) \lor A(\f{p}) = \f{NT}) \\
 \end{cases}
$$
where $T$ is the set of available test cases (both passing and failing tests), 
$\f{p} \vdash t$ indicates that the examined patch $\f{p}$ runs successfully against the test $t$, and
$A$ is a termination prover.  

\subsection{Monotonicity of Loop Programs}

\noindent
We now turn to discuss the class of monotonic statements~\cite{Gupta, Madalene} that is often encountered in loop programs. 
The monotonicity of a statement is defined with respect to a specific loop surrounding the statement.
Consider while loop $P$ and a statement $s : x = e$ inside the loop. 
Further, consider a single execution of the loop, which involves $n$ iterations through the loop. 
Let $\ell_1, \ell_2,...,\ell_n$ denote the $n$ consecutive iterations of the loop,
and $x_1, x_2,.., x_m$ denote the values assigned to $x$ during these iterations, 
where $m \leq n,$ because statement $s$ may not be executed during every iteration if there are conditional branches inside the loop.

\begin{definition} \label{MonotonicProperty} \textbf{Monotonic Loop Statements}. %
A statement $s : x = e$ is considered to be \emph{monotonic} w.r.t. loop $P$ iff the sequence of values assigned to variable $x$
during successive executions of $s$ forms an increasing or decreasing sequence of values (
i.e. $x_i < x_{i+1}$ or $x_i > x_{i+1}$).
A statement $s$ is considered to be \emph{regular monotonic} iff the sequence $x_1, x_2,.., x_m$ is an arithmetic progression or geometric progression;
it is considered to be \emph{irregular monotonic} otherwise.
\end{definition}

The monotonicity of a statement $s : x = e$ w.r.t. loop $P$ can be determined using various approaches. 
Spezialetti and Gupta present a sophisticated static analysis technique to determine loop monotonic variables~\cite{Madalene}.
Alternatively, one can verify whether the given loop program meets the monotonicity property by executing the program $P$ against the available test cases, 
and checking whether the values assigned to the control variables follow a monotonic function cf. Definition~\ref{MonotonicProperty}. 
The key challenge is then to synthesize monotonic update expressions for control variables that ensure proper termination of the buggy loop program under repair.
We show in Section~\ref{sec:implementation} how one can exploit the monotonicity property to guide the repair algorithm toward feasible patches.

\subsection{Repair Procedure for Termination Bugs}

\noindent
The most straightforward approach to search-based program repair uses random mutation of expressions and program statements to generate candidate patches.
Observe that the search space for potential patches in this approach can be extremely large, even for programs whose source code size is small,
as each statement in the program can be mutated using different mutation operators, such as insert, delete, and replace, 
and with different parameters.
As a result, the number of potential mutants grows exponentially w.r.t. the number of lines in the code and measures need to be taken to 
ensure the efficiency and performance while examining candidate patches in the generated search space. 

However, instead of randomly mutating statements of a given buggy non-termination loop program $P$, 
one can direct or guide the genetic repair algorithm to focus on the set of (feasible) statements whose mutation may lead to valid repairs for termination bugs. 
The set of feasible statements for mutation would be the set of statements that directly or indirectly affect the evaluation of the termination condition of the given buggy program.

Program slicing is a viable technique to restrict the focus of the repair task to specific parts of a program~\cite{Weiser81}.
It has been applied successfully in software engineering tasks,
including debugging, testing, program restructuring, and downsizing. 
We apply program slicing in APR so that statements that are not relevant to the detected bugs may be skipped when searching for repairs.
Among the various existing slicing algorithms, 
we choose a generalized slicing paradigm, called ``conditioned'' slicing~\cite{Harman2001}.
It is a generalization of static slicing and dynamic slicing.
A conditioned slice is constructed for a slicing criterion that includes the condition which causes the program to misbehave.
To compute the set of feasible statements for mutation, 
we need first to compute the set of control variables that affect the termination of $P$ 
(i.e., the set of variables whose values determine the end of the loop). 

\begin{definition} \textbf{Loop Control Variables}.
Let $P$ be a loop program and $X$ be the set of variables of $P$. 
Let also $\mathcal{C}(P)$ be the termination expression of $P$ 
(i.e., a logical expression whose evaluation determines whether or not the loop $P$ will iterate). 
We refer to $x \in X$ as a \emph{control variable} of $P$ iff the value of $x$ affects the truth value of $\mathcal{C}(P)$.
\end{definition}

With these elements in place, we can formulate an algorithm for generating valid repairs for non-terminating loop programs.
Let $P_{\f{min}}$ be a minimized loop program of a given buggy non-terminating loop program $P$ that contains only the set of statements that affect the termination of $P$. 
$P_{\f{min}},$ is constructed from $P$ by applying a slicing algorithm in which the set of control variables are used as slicing points. 
By mutating the statements of $P_{\f{min}}$, we construct a search space of patches for the detected termination bug (a.k.a. ``patch space'').
A reliable termination prover $A$ together with the available test cases $T$ are used to validate generated patches cf. Definition~\ref{ValidityDef}. 
The key algorithmic steps to generate a valid repair for $P$ can then be summarized as follows:
\begin{enumerate}
 \item Compute the set of control variables of the loop $P$.
 \item Construct a minimized loop program $P'$ from $P$ by using control variables of $P$ as slicing criteria. 
 \item If $P'$ meets the monotonicity property, then construct the candidate patch space $S$ 
    by monotonic mutation of update expressions of control variables. 
    Otherwise, use a genetic repair approach similar to GenProg.
 \item Select a patch $p$ from the constructed patch space $S$. 
 \item Use a termination prover together with available test cases to check the validity of $p$.
 \item If $p$ is a valid patch or the allocated time budget is expired, then return. Otherwise, go to step 4.
\end{enumerate}
The first three steps of the algorithm can be performed by employing static analysis techniques together with slicing techniques. 
The goal of these steps is to reduce the search space of feasible patches, in which a correct repair can be found faster. 
The algorithm terminates if either the allocated time budget is expired or a valid patch is found.
Observe that the soundness of generated patches for termination bugs will be relative to the soundness of the termination prover that is employed.
Moreover, note that mutation of the termination expressions and update expressions of the control variables for a loop program $P$ 
does not only affect the termination of $P$, but likely also affects the functionality of $P$.

\section{Initial Prototype and Analysis}%
\label{sec:implementation}

\noindent
In our initial prototype, we focus mainly on the repair of termination bugs, 
as this class of bugs has received significantly less attention in the APR 
literature.\footnote{~Note that the discussion in Section~\ref{subsection:reliability} includes the repair of IO bugs.}
As far as we know, there is no available dataset for termination bugs in loop programs. 
To extract some useful information about common syntactic shapes of both the update expressions of control variables and termination expressions of successfully terminating loop programs, 
we perform an initial analysis on the available loop programs in the following two datasets. 
\begin{enumerate}
\item The SNU real-time benchmark suite containing small C programs for worst-case execution time analysis.\footnote{~Available at \url{www.cprover.org/goto-cc/examples/snu.html}}
\item The Power-Stone benchmark suite as an example set of C programs for embedded systems~\cite{Ku2007}. 
\end{enumerate}
These datasets have been constructed to compare the efficiency and reliability of different available termination provers.
Analyzing the two datasets with the termination provers AProVE and 2LS allows us to make the following key observations:
\begin{itemize}
 \item The tool 2LS was able to prove termination by returning definite answers for almost $80\%$ of the examined loop programs,
    while AProVE was able to prove termination for only $37\%$ of the programs.
    However, there were few cases (around $3\%$ of the examined programs) 
    where AProVE was able to prove termination while 2LS not.
    The two tools together were able to prove the termination of around $84\%$ of the examined programs.
    Termination provers return ``unknown'' when they are unable to prove termination of the program. 
    This often occurs due to the high complexity of the loop program under analysis.
 \item We did not identify any cases in which the two tools returned contradicting answers for the same loop program 
    (i.e., where one tool answered ``terminating'' and the other answered ``non-terminating'').
    This increases our confidence about the soundness of implemented theories in both tools.
 \item The initial exploration shows that the termination provers AProVE and 2LS are able to verify termination of programs using very little computational time (a few seconds). 
    We also observe that it is more convenient to use the termination provers directly, 
    without extracting termination rules for the loop program under repair. 
\end{itemize}
The question is then which tool to use for integration when fixing termination bugs: 2LS or AProVE? 
The analysis conducted on the two datasets showed that 2LS was able to prove termination for a larger number of loop programs than AProVE.
However, this observation may vary depending on the complexity of the program under analysis. 
It is more beneficial to consider both tools when validating generated patches for termination bugs. 
That is, we may run both tools in parallel against the examined loop program. 
To reduce the amount of overhead introduced by the tools, one may choose to run the tools only against plausibly generated patches 
(i.e., patches that successfully passed available test cases). 

\subsection{Monotonicity Property in Practice}

\noindent
For this analysis, we consider the subset of successfully terminating loop programs in the two datasets, as verified by termination provers. 
We study the syntactic shapes of both the termination condition and the update expressions of control variables for each loop program. 
The goal is to infer common patterns that can be used to guide the repair engine to generate valid patches for termination bugs.

Our analysis shows that the SNU suite contains 107 loop programs, and 105 of them have monotonic behavior. 
We also observe that the update expressions for 63 loop programs in SNU are simple monotonic expressions 
(i.e., $u(x) = x~ op ~b$, where $op \in \{+, -, *, \div\}$ and $b$ is a constant). 
The Power-Stone benchmark suite contains 112 loop programs, of which 110 have monotonic behavior. 
This implies that $98\%$ of the loop programs in both suites are monotonic programs.

We classify the variables in termination expression $\mathcal{C}(P)$ based on their boundedness into variables that are bounded from below and variables that are bounded from above:
\begin{definition} \label{boundedVariables} \textbf{Bounded Control Variables}.
Let $P$ be a loop program and $X$ be the set of control variables of $P$ and $\mathcal{C}(P)$ be the termination expression of the loop $P$.
We say that a variable $x_i \in X$ is bounded from below in $\mathcal{C}(P)$ if it has the form $(x_i \sim_i c_i)$, where $\sim_i \in \{>, \geq\}$ and $c_i$ is called the lower bound of $x_i$. 
On the other hand, we say that a variable $x_j \in X$ is bounded from above in $\mathcal{C}(P)$ if it has the form $(x_j \sim_j c_j)$, where $\sim_j \in \{<, \leq\}$ and $c_j$ is called the upper bound of $x_j$.
\end{definition}

Moreover, we classify the update expressions for control variables into monotonically increasing expressions and monotonically decreasing expressions cf. Definition~\ref{MonotonicProperty}.
Such classification of control variables in termination conditions and update expressions is crucial, 
as it determines the feasible direction of mutation of logical expressions and arithmetic expressions of the loop program under repair.

Observe that for the feasibility of generated patches for termination bugs, 
the termination expression $\mathcal{C}(P)$ needs to be mutated while considering the syntactic shape of the update expression $u(x_i)\mid x_i \in X$ and vice versa.
This is necessary in order to detect early infeasible candidate patches.
In fact, the shape of the expression $\mathcal{C}(P)$ imposes some restrictions on the mutation function that is used for the update expressions of control variables of the loop program under repair.
This leads to the notion of conditional mutation of expressions. 
\begin{definition} \textbf{Conditional Mutation of Expressions}.
Conditional mutation is an operation where the mutation of some expression $e_i$ in a program $P$ depends on the syntactic shape of a related expression $e_j$.
For a loop program $P$ with a set of control variables $X$, 
the mutation of $\mathcal{C}(P)$ depends on the syntactic shape of $u(x_i)$ and vice versa.
\end{definition}

Note that both expressions $\mathcal{C}(P)$ and $u(x_i)$ affect termination of the loop $P$ 
and that the syntactic shape of $u(x_i)$ affects the evaluation of $\mathcal{C}(P)$ between the successive iterations of the loop.
To spend the available time budget in a more efficient manner, 
conditional mutations play a crucial role, 
as they allow detecting and skipping infeasible patches even before constructing and validating them.
To develop a better understanding of conditional mutation, let us consider the following example:
\begin{example}
Consider a loop program $P$ with a single control variable $x$. 
Let $\mathcal{C}(P) = (x \sim c)$ and $u(x) = (x~ op ~b)$,
where $op$ is an arithmetic operator and $\sim$ is a comparison operator. 
It is easy to see that the mutation of the operator $op$ will affect the evaluation of $\mathcal{C}(P)$ 
and hence the mutation choices of $op$ should be made while taking into account the operator $\sim$. 
For instance, for the case where $\sim = `<'$ and $b > 0$ the mutations of operator $op$ should not consider the two operators $\{-, \div \}$ because $x$ is bounded from above in $\mathcal{C}(P)$, and mutating $op$ with $-$, or $\div$ would lead to a monotonically decreasing expression, which would lead to invalid patches. 
\end{example}

Let $\f{monotoneMutate}{}(e)$ be a mutation function that takes some monotone expression $e$ and produces another new monotone expression $e'$. 
We assume that the function $\f{monotoneMutate}{}(e)$ implements some static analysis techniques to check the monotonicity of expressions,
similar to those implemented by Spezialetti and Gupta~\cite{Madalene}.
Analysis of successfully terminating programs in the two datasets yields a number of useful conditional mutation rules:
\begin{itemize}
 \item If the condition $\mathcal{C}(P)$ or some sub-condition in $\mathcal{C}(P)$ has the form $(x \sim c)$ where $\sim \in \{>, \geq\}$, then the update expression $e = u(x)$ needs to be mutated by the function $\f{monotoneMutate}{}(e)$ such that the resultant expression $u'(x)$ is a monotonically decreasing expression. 
 \item If the condition $\mathcal{C}(P)$ or some sub-condition in $\mathcal{C}(P)$ has the form $(x \sim c)$ where $\sim \in \{<, \leq\}$, then the update expression $e = u(x)$ needs to be mutated by the function $\f{monotoneMutate}{}(e)$ such that the resultant expression $u'(x)$ is a monotonically increasing expression. 
 \item If $u(x)$ has the form $(x ~ op ~b)$ where $op \in \{ +, *\}$ and $b > 0$ and $x \in \mathit{R^{+}}$,
 then mutate $\sim$ in the expression $(x \sim c)$ using the set of operators $\{>, \geq , =\}$. On the other hand, if $op \in \{ -, \div\}$ and $b > 0$ and $x \in \mathit{R^{+}}$
 then mutate $\sim$ in $(x \sim c)$ using the set of operators $\{<, \leq , =\}$.
\end{itemize}
One can develop several similar conditional mutation rules by exploiting the monotonicity property of loop programs 
and the boundedness direction of control variables in termination conditions.
Note that the same expression $e$ can be mutated to be monotonically increasing or monotonically decreasing expression, 
depending on how we mutate the ingredients of the expression.
It is easy to see that by following the above conditional mutation rules when mutating non-terminating loops (whenever applicable), 
we guarantee that there will be an iteration at which the termination condition of the loop will be evaluated to $\f{false}$ and the loop terminates.

\subsection{Reliability of Repair Approaches}\label{subsection:reliability}

\newcommand{\MANYBUGS}{\textsc{manybugs}}
\newcommand{\INTROCLASS}{\textsc{introclass}}

\noindent
To verify the reliability of existing test-based repair approaches in generating valid patches for IO bugs and termination bugs, 
we consider several datasets that have been used by the tools GenProg and SCRepair. 
The datasets contain a considerable number of programs that suffer from IO bugs and termination bugs. 
The dataset used by the SCRepair tool contains three programs with 12 IO bugs. 
The datasets used by GenProg, namely the \MANYBUGS{} and \INTROCLASS{} benchmarks, 
contain in total 1,183 different bugs or defects associated with test cases, spread over 15 C programs.

It has been claimed that GenProg can repair many kinds of defects, 
including non-terminating loops and integer overflows, 
based on the observation that the tool can generate plausible patches for most of these classes of bugs. 
However, the validation process is performed using weak specifications for both IO bugs and termination bugs, 
due to the assumption that accurate, complete specifications are typically unavailable. 
In fact, one can synthesize accurate specifications for these particular classes of bugs 
by taking advantage of available reliable bug detection tools, as demonstrated in this paper.

We examine the set of plausible patches generated by GenProg and SCRepair for the available buggy IO programs and non-terminating loop programs, while considering the correctness properties introduced in this work to verify their soundness.
The analysis shows that none of the plausibly generated patches for both IO bugs and termination bugs were correct. 
That is, none of the generated patches for IO bugs meets specification $\f{Spec}_{\f{IO}}$
(i.e., the correctness specification for IO bugs), 
and none of the generated patches for buggy non-terminating loops successfully passes the validation process performed by the termination provers.
We also observe that the repair tools do not consider a composite correctness property when validating generated patches for termination bugs (i.e., non-terminating loops). 
This raises questions about the reliability of test-based repair approaches that do not use special mutation functions that take the semantics of the bug being repaired into account.

Note that we do not consider a patch that fixes an IO bug \emph{valid} if the patch breaks the intended semantics of the original buggy arithmetic expression.
Similarly, we do not consider a patch that ensures termination of a given non-terminating loop program valid if the patch breaks the intended semantics of the loop program. 

To improve the reliability of APR, innovative approaches are required for both the patch generation and patch validation steps.
Search-based repair approaches can be promising candidates, 
provided that the buggy program is mutated using special mutation functions that take the semantics of the bug into account, 
and provided that patch validation is performed using accurate specifications that are synthesized from the knowledge contained in static bug detection rules.

\section{Related Work}
\label{sec:relwork}

\noindent
We distinguish the following categories of related work:

\head{Automated Program Repair}
We limit ourselves to offline, source-based, automated program repair approaches~\cite{monperrus2018:automatic}.
These can be separated into two classes: search-based approaches and semantic-based approaches. 
Search-based approaches such as Genprog~\cite{GenProg}, Astor~\cite{martinez2016:astor}, 
and SCRepair~\cite{TOSEM2020} predominantly use failing test cases to identify bugs, 
and then apply mutations to the source code until the program passes all failing test cases.
These approaches do not provide patch correctness guarantees beyond the fact that the provided test cases now pass. 
Furthermore, these approaches require executing the buggy program, 
first to find the bug in the program, and then to generate and validate candidate repairs. 
Semantic-based approaches like SemFix~\cite{Nguyen}, Nopol~\cite{Xuan}, DirectFix~\cite{Mechtaev}, 
SPR~\cite{Long15stagedprogram}, Angelix~\cite{Mechtaev16}, and JFIX~\cite{Xuan2017} 
infer repair constraints for the buggy program via symbolic execution of the given tests. 
The completeness of inferred repair constraints relies on the size and quality of the available test-suite.

\head{Detecting IO bugs}
There have been a number of approaches developed to detect integer overflow at the source code level. 
These approaches can be classified into two categories: 
(a) instrumenting the source code with run-time integer overflow check~\cite{BrumleySCJL07,ZhangSDCZFF15,DietzLRA15},
and (b) using static analysis to detect integer overflow~\cite{Coker,LogozzoM13,Wang2012}. 
Of these, the work of Coker and Hafiz~\cite{Coker} comes closest to the work presented here, 
by introducing a set of refactoring and rewrite rules to apply in an IDE to fix overflows in C programs. 
However, unlike the work presented in this paper, 
they do not propose any way to automatically generate fixes and verify them.

\head{Evaluating Overfitting in APR}
A number of studies have evaluated overfitting in APR, 
and the overall outcome of these studies is that the accuracy of test-based repairs is too low~\cite{QiLAR15, MartinezDSXM17}. 
A study conducted by Qi et al.~\cite{QiLAR15} shows that GenProg~\cite{GenProg}, 
one of the most well-known program repair techniques, 
produced plausible patches for 55 different defects,
but only two were correct, giving a precision of $4\%$.
Yang et al.~\cite{YangZLT17} propose a technique to improve evaluation of the correctness of generated plausible patches 
by extending the number of test cases in a test suite using the fuzzer American Fuzzy Lop (AFL). 
Their study identified 321 overfitted patches out of 427 examined plausible patches generated by GenProg,
Kali~\cite{QiLAR15}, and SPR~\cite{SPR}. 

Le et al.~\cite{Le} examine different ways to measure overfitting in dynamic APR approaches, 
using independent tests and manual inspection of patches. 
They conclude that neither human judgment nor independent testing can truly determine overfitting.
Ye et al.~\cite{Ye} evaluated five repair systems based on the QuixBugs benchmark~\cite{LinKCS17} consisting of 40 small-sized Java buggy programs. 
Their results show that 64 patches were generated for 15 individual programs. 
They evaluated the correctness of the patches by generating more tests using EvoSuite~\cite{FraserA11}, 
as well as via manual analysis. 
Their analysis shows that 33 out of 64 generated patches were overfitting.

The above-described evaluation studies relied mainly on incomplete or insufficient specifications, and hence they may not discover all existing overfitted patches.
Instead of evaluating plausible patches by increasing the number of tests in the test suite or by manual inspection of the code, it is highly desirable
to synthesize complete and accurate specifications for recurring classes of bugs by utilizing knowledge contained in static bug detection tools as we have done in this work. 

\head{Alternative Specification Sources for APR}
Several attempts have been made to use other sources of information than test suites to formulate correctness specifications for APR.
Examples include pre- and post-conditions, abstract behavioral models specified by the user, 
and the application of static analysis tools as oracles~\cite[\S3.1-3.4]{monperrus2018:automatic}.
Of these, the application of static analysis tools as oracles comes closest to the work proposed here, 
but as discussed in Section~\ref{sec:integrate}, such a direct integration has a number of disadvantages.  
Note that none of these approaches use the actual static detection patterns/rules as the source for formulating accurate specifications.
There also exists a few examples of using information from debugging to aid APR:
Facebook's APR tool SapFix takes information generated during the bug detection process and applies various techniques, 
including a template-based one, specific to a given bug, to fix the program. 
However, our approach is different in that we add accurate specifications to APR to check for overfitting patches. 
Moreover, it considers new classes of bugs whose fixing requires the satisfaction of composite properties.

\section{Concluding Remarks}
\label{sec:conc}

\subsection{Contributions and Key Findings}
\noindent
In this paper, we study the feasibility of integrating static bug detection and automated program repair so that repairs may be generated in a faster and more reliable manner.
The feasibility is examined for two classes of bugs: arithmetic bugs, such as IO bugs, and logical bugs, such as termination bugs.
Fixing IO bugs has been studied before, but with weaker specifications 
(mainly specifications that are synthesized based on test cases), 
which may not guarantee the correctness of generated patches. 
Termination bugs have not been studied in great detail in the prior literature. 
To our knowledge, this is the first work that synthesizes complete specifications for these classes of bugs. 

\smallskip\noindent
The key findings of this work can be summarized as follows:
\begin{itemize}
 \item General-purpose APR tools treat different classes of bugs in the same manner: 
  the repair algorithms implemented by these tools do not take the specific characteristics of the bug being repaired into account. 
  Experiments with GenProg and SCRepair on IO and termination bugs show that none of the plausible patches generated by these tools are correct. 
  \emph{Integration of static bug detection in the repair process helps to reduce the search space and improve the reliability of APR}.

 \item Pattern-based formal specifications are more reliable as the oracle for correct behavior than test-based specifications. 
  The key distinguishing feature is that pattern-based specifications are more general and therefore provide broader coverage of programs that can be repaired. 
   
 \item Patch validation of IO bugs is more complex than one might initially anticipate. 
  In fact, the complexity of the validation process varies depending on the patching technique that is applied. 
  If expression rewriting rules are used to generate patch candidates, 
  these need to be \emph{validated using a composite correctness property} 
  that checks absence of overflows and preserving the original semantics (cf.\ $\f{Spec}_{\f{IO}}$ in Eq.~\ref{spec_io}). 
  On the other hand, if variable widening is used to generate patch candidates, 
  then all \emph{dependent} variables in the program need to be widened as well, 
  and validation needs to check all arithmetic expressions to which the widened variables \emph{contribute}.
  Such a complex patch validation process is required to ensure that no new bugs are introduced when widening some variables in the program under repair.

 \item Special mutation functions should be synthesized for different classes of bugs, depending on the semantics of the bugs. 
  For example, for IO/IU bugs, the mutation function should be designed in a way such that the generated expression is semantically equivalent to the original buggy expression. 
  For termination bugs, it is desirable to generate monotonic expressions for update expressions of the control variables, 
  and the mutation should take the syntactic shape of the termination expression into account (i.e., using conditional mutation).
\end{itemize}

\subsection{Directions for Future Work}

\noindent
In this initial exploration, we focus mainly on the problem of \emph{patch validation} rather than \emph{patch generation} in APR. 
To complete the line of research initiated here, we identify the following directions for future work.

First and foremost, additional case studies need to be performed to analyze and demonstrate the feasibility and limitations of the proposed approach on a wider range of bugs.  

Next, a (semi-)automated mining technique needs to be devised that can derive bug detection patterns for various classes of bugs from a selection of reliable static analysis tools. 
Our initial focus will be on mining patterns from static analysis tools with configurable and customizable bug detection rules.

Subsequently, novel patch generation procedures need to be implemented for arithmetic bugs and termination bugs that exploit the mined 
detection patterns as correctness specifications to reduce the search space and increase reliability.

Finally, the efficacy of the proposed approach needs to be evaluated. 
To this end, comprehensive datasets need to be constructed and curated, in particular for loop programs with termination bugs where the existing datasets were made for a different purpose and may be suboptimal for evaluating APR.

\balance
\printbibliography 

\end{document}